\documentclass[%
reprint,
superscriptaddress,
nofootinbib,
amsmath,
amssymb,
aps,
pra,
iopjournal
]{revtex4-2}

\usepackage{graphicx}
\usepackage{dcolumn}
\usepackage{bm}
\usepackage{orcidlink}
\usepackage{amsthm}
\usepackage{hyperref} 

\usepackage{mathtools}
\usepackage{relsize}
\usepackage{csquotes}
\usepackage{multirow}
\usepackage{booktabs}
\usepackage{bbm}
\usepackage{placeins}
\usepackage{tikz}
\usetikzlibrary{quantikz2}
\usepackage{chemfig}
\usepackage[framemethod=TikZ]{mdframed}
\usepackage{tabularx,ragged2e}
\usepackage{float}

\hypersetup{
    colorlinks=true,
    linkcolor=blue,
    filecolor=magenta,      
    urlcolor=blue,
    citecolor=blue
}

\newtheorem*{theorem}{Theorem}
\newtheorem*{property}{Property}

\newtheorem*{corollary}{Corollary}

\definecolor{bgblue}{RGB}{245,243,253}
\definecolor{ttblue}{RGB}{91,194,224}
\definecolor{deepblue}{rgb}{0,0,0.5}
\definecolor{deepred}{rgb}{0.6,0,0}
\definecolor{deepgreen}{rgb}{0,0.5,0}
\definecolor{deepgreen}{HTML}{395B64}
\definecolor{darkgreen}{HTML}{2C3333}
\definecolor{grey}{HTML}{808080}

\begin{document}

\title{Encoding molecular structures in quantum machine learning}

\author{Choy~Boy\,\orcidlink{0000-0002-6782-9433}}
\thanks{Equal contributions, shared first-authorship.}
\affiliation{Yusuf Hamied Department of Chemistry, University of Cambridge, Lensfield Road, Cambridge CB2 1EW, United Kingdom}
\affiliation{The Hartree Centre, STFC, Sci-Tech Daresbury, Warrington, WA4 4AD, United Kingdom}

\author{Edoardo~Altamura\,\orcidlink{0000-0001-6973-1897}}
\thanks{Equal contributions, shared first-authorship.}
\affiliation{The Hartree Centre, STFC, Sci-Tech Daresbury, Warrington, WA4 4AD, United Kingdom}
\affiliation{Yusuf Hamied Department of Chemistry, University of Cambridge, Lensfield Road, Cambridge CB2 1EW, United Kingdom}

\author{Dilhan~Manawadu\,\orcidlink{0000-0002-3575-8060}}
\affiliation{The Hartree Centre, STFC, Sci-Tech Daresbury, Warrington, WA4 4AD, United Kingdom}

\author{Ivano~Tavernelli\,\orcidlink{0000-0001-5690-1981}}
\affiliation{IBM Quantum, IBM Research Zurich, 8803 R{\"u}schlikon, Switzerland}

\author{Stefano~Mensa\,\orcidlink{0000-0002-0938-144X}}
\affiliation{The Hartree Centre, STFC, Sci-Tech Daresbury, Warrington, WA4 4AD, United Kingdom}

\author{David~J.~Wales\,\orcidlink{0000-0002-3555-6645}}
\thanks{Contact author: \href{mailto:dw34@cam.ac.uk}{dw34@cam.ac.uk}}
\affiliation{Yusuf Hamied Department of Chemistry, University of Cambridge, Lensfield Road, Cambridge CB2 1EW, United Kingdom}

\date{\today}

\begin{abstract}

Quantum machine learning (QML) has  great potential for the analysis of chemical datasets. However, conventional quantum data-encoding schemes, such as fingerprint encoding, are generally unfeasible for the accurate representation of chemical moieties in such datasets. In this contribution, we introduce the quantum molecular structure encoding (QMSE) scheme, which encodes the molecular bond orders and interatomic couplings expressed as a hybrid Coulomb–adjacency matrix, directly as one- and two-qubit rotations within parameterised circuits. We show that this strategy provides an efficient and interpretable method in improving state separability between encoded molecules compared to other fingerprint encoding methods, which is especially crucial for the success in preparing feature maps in QML workflows. To benchmark our method, we train a parameterised ansatz on molecular datasets to perform classification of state phases and regression on boiling points, demonstrating the competitive trainability and generalisation capabilities of QMSE. We further prove a fidelity-preserving chain-contraction theorem that reuses common substructures to cut qubit counts, with an application to long-chain fatty acids. We expect this scalable and interpretable encoding framework to greatly pave the way for practical QML applications of molecular datasets.

\end{abstract}

\maketitle

\section{Introduction}
\label{sec:intro}

The integration of machine learning (ML) techniques  in chemistry has led to significant advances, such as improved prediction of protein structures \cite{jumper2021highly} and the estimation of blood-brain barrier permeability for small molecules as potential drug candidates \cite{huang2024predicting}. Quantum computing is a promising approach for  enhancing ML pipelines involving classical and quantum data \cite{biamonte2017quantum}. In this context, quantum machine learning (QML) algorithms have been proposed to improve \textit{in-silico} screening and quantum‐assisted drug design \cite{Mensa_2023,li2024hybrid,sundaram2025challenges}. In the near term, QML may continue to deliver practical advantages in specialised tasks, such as learning from quantum data, simulating physical systems, and employing quantum feature spaces, particularly when paired with hybrid quantum-classical architectures. As quantum hardware is set to improve with longer qubit coherence times \cite{PhysRevLett.130.267001, Tuokkola2025NearMS, Réglade2024QuantumControl}, reduced leakage \cite{miao2023overcoming}, and suppressed cross-talk \cite{PhysRevApplied.18.024068}, QML models may outperform their classical counterparts in representing and optimising high-dimensional, structured, and entangled data, especially in domains like quantum chemistry, material science, and drug discovery. This potential is expected to become even more significant in the fault-tolerant quantum computing (FTQC) regime, where QML is expected to offer competitive speedups for variational \cite{cong2019quantum} and kernel methods, feature selection \cite{yamasaki2020learning}, and generative models \cite{hibat2024framework, riofrio2024characterization}. In addition, various studies involving QML have found significant advantages via the combination of superior generalisability \cite{gil2024understanding} alongside higher accuracies for less training data inventories \cite{gupta2022comparative} compared to their classical equivalent. As both software and hardware frameworks continue to advance, QML is poised to become a foundational element in achieving quantum advantage across computational learning and scientific discovery \cite{PRXQuantum.3.030101}; even more so when next-generation fault-tolerant quantum devices and algorithms  become available.

Besides trainable architectures, a key component of any ML pipeline is the choice of encoding data as input vectors. In particular, molecular representation learning seeks to optimise the transformation of molecular structures as suitable input vectors in trainable models \cite{boulougouri2024molecular}. Standard techniques, such as one‐hot encoding \cite{potdar2017comparative} and its embedded variants \cite{nunez2025embedded} are effective in partially alleviating the \lq curse of dimensionality' associated with representing molecular structures by compressing high‐dimensional binary vectors into lower‐dimensional real‐valued representations. In addition, more sophisticated techniques such as graph-based molecular representation learning methods (e.g. group graphs \cite{cao2024group}) improve upon atom-level encodings by representing substructures as nodes and encoding connectivity via edges.

Quantum encoding schemes, such as basis encoding, angle encoding, and amplitude encoding, map classical features into data-encoding quantum circuits for QML processes~\cite{smaldone2025}. Basis encoding typically represents a binary molecular fingerprint of length $\tau$ directly into $\tau$ qubits, but this procedure becomes unfeasible for larger fingerprints. Amplitude encoding reduces qubit requirements to $O(\log \tau)$ by mapping a normalised feature vector into the amplitudes of a quantum state, but preparing arbitrary amplitude‐encoded states requires an exponential scaling of two‐qubit gates in the worst case \cite{Mottonen2005,Araujo2021,Shende2006}, rendering it impractical on near‐term devices. Angle encoding, which parameterises one‐qubit rotations with feature values, offers a hardware‐efficient alternative, but can suffer from poor state separation and trainability issues, especially when paired with dimension-reduced features to reduce quantum hardware requirements. These limitations motivate  development of new encoding techniques that strike a balance between expressivity, trainability, and resource efficiency.

{There is growing traction in the number of strategies devised for the deployment of quantum neural networks in predicting chemical phenomena, in particular the training and prediction of potential energy surfaces and molecular force fields \cite{Xia2020, Kiss2022, Le2025}.} Recent studies have also proposed techniques to encode molecular properties in Hilbert spaces accessible via quantum computing. \textcite{boiko2025advancing} introduced stereoelectronics-infused molecular graphs (SIMGs), which enrich traditional molecular graphs by incorporating orbital-centric nodes (e.g., $\sigma$, $\pi$, $\sigma^*$, $\pi^*$, lone pairs) and quantified donor–acceptor interactions derived from Natural Bond Orbital analysis~\cite{weinhold2001natural}. A surrogate graph neural network is trained to predict these features directly from 3D molecular geometries, enabling fast and accurate inference for downstream property prediction. This approach enhances model interpretability and generalises to large biomolecules. Compared to classical descriptors, namely Coulomb matrices~\cite{rupp2012fast}, SOAP~\cite{bartok2013representing}, or graph-based encodings such as \texttt{ChemProp}~\cite{heid2023chemprop}, surrogate graph neural networks offer superior chemical fidelity by encoding quantum interactions explicitly. The advantages of such models include interpretability and high performance in message-passing neural networks, while their limitations include the initial computational overhead of quantum chemical calculations and the need for dataset-specific retraining when the datasets are extended.

\textcite{torabian2025molecular} proposed a novel isomorphism between quantum circuits and polyatomic molecules, enabling the mapping of circuit architectures to molecular descriptors, such as Coulomb matrices, molecular fingerprints, and Gershgorin discs. These descriptors can be used to predict the performance of quantum support vector machines, offering a strategy to reduce the search space in circuit design. Their method complements efforts to mitigate barren plateaux~\cite{larocca2025barren}, exponential kernel concentration~\cite{thanasilp2024exponential}, and noise-induced degradation in kernel methods~\cite{huang2024predicting}. They also relate to advances in covariant \cite{glick2024covariant} and Fisher kernels that aim to preserve relevant data structure. The main advantage is the physically interpretable restriction of circuit composition using descriptors well-established in chemoinformatics. However, limitations include the potential ambiguity in reverse-mapping molecules back to unique quantum circuits and scalability concerns for deeply layered architectures or high-qubit-count regimes.

Finally, \textcite{kamata2025molecular} developed the molecular quantum transformer (MQT), a hybrid classical–quantum architecture that uses quantum self-attention to represent and predict molecular ground-state energies. The model encodes bond length–dependent molecular Hamiltonians via parameterised quantum circuits and exploits training on multiple geometries for efficient learning of potential energy surfaces. In contrast to methods like Variational Quantum Eigensolvers (VQE)~\cite{peruzzo2014variational} or meta-VQE~\cite{cervera2021meta}, which require separate circuit evaluations for each molecular configuration, MQT offers a more data-efficient alternative. It also outperforms classical Transformer models when learning from small datasets and supports pretraining and fine-tuning workflows. However, its reliance on amplitude encoding and large circuit ans\"atze may limit feasibility on near-term hardware. The work aligns with recent proposals for quantum-enhanced transformers~\cite{li2024quantum} and builds on advances in neural-network quantum states and denoising~\cite{tran2020gan}.

Despite early theoretical and experimental advances \cite{liu2021rigorous}, realising quantum advantage in supervised QML remains challenging in training strategies like quantum neural networks (QNNs) due to barren plateaux, high circuit depth, noise, and the expressivity of parameterised quantum circuits \cite{liu2021rigorous,thanasilp2023subtleties, thanasilp2024exponential,larocca2025barren, Crognaletti2025}. Other works have shown that trainable circuit architectures, such as convolutional QNNs, can be classically simulated \cite{bermejo2024quantum}, attenuating potential advantages of using quantum devices as opposed to classical computers. In some cases, such prospects for exponential speedups over classical ML are found to be generated by explicit or implicit assumptions introduced in mathematical proofs \cite{PhysRevLett.127.060503}, making practical quantum advantage uncertain \cite{PhysRevLett.126.190505}.

In this work, we introduce the \emph{quantum molecular structure encoding} (QMSE) scheme, which explicitly encodes molecular bond orders and interatomic couplings via a hybrid Coulomb–adjacency matrix as parameterised one‐ and two‐qubit rotation gates in the data-encoding quantum circuit in QML workflows. This approach addresses several key challenges identified in recent QML literature. First, rigorous quantum speed‐up results for supervised learning tasks suggest that specialised feature maps can yield provable advantages \cite{PRXQuantum.3.030101,liu2021rigorous}, but only if they produce sufficiently distinct quantum states; QMSE’s graph-based representation can achieve a broader distribution of fidelities compared to conventional fingerprint (angle) encoding. Second, subtleties in trainability and barren plateau effects have been shown to impede variational QML models \cite{thanasilp2023subtleties}; by constructing an encoding that exploits commutativity of two‐qubit interactions (e.g., $R_{xx}$ rotations), QMSE provides a more robust optimisation landscape. Third, exponential concentration in quantum kernel methods can render quantum-enhanced similarity measures ineffective for high-dimensional classical data \cite{thanasilp2024exponential}; {the one-to-one physical mapping of the QMSE scheme allows for a more bespoke representation of molecular datasets as data-encoding circuits that better reflect chemical similarity, and thus we expect QMSE to alleviate the saturation issues associated with traditional fingerprint-based kernels.} Finally, the burgeoning success of classical large language models in few-shot learning \cite{brown2020language} underscores the importance of scalable, data-efficient architectures. QMSE draws inspiration from this approach by encoding molecular graphs in a structured, modular fashion, thereby facilitating generalisation in small-dataset regimes typical of chemical screening.

Compared to graph‐based classical molecular representation learning, QMSE directly incorporates quantum‐chemical insights, such as bond orders, interatomic couplings, and stereochemistry, into the single and entangling quantum gates of circuits in the data-encoding layer of QML, akin to a quantum approximate optimisation algorithm (QAOA) circuit for connected graphs. We demonstrate that QMSE not only reduces resource demands on near-term quantum hardware, but also yields significantly higher training and test accuracies, outperforming standard fingerprint encoding in both classification and regression tasks on chemical datasets. Furthermore, we prove a fidelity‐preserving chain‐contraction theorem that eliminates common molecular fragments in reducing qubit counts, paving the way for scalable QML applications to long‐chain molecules and large datasets.

This paper is organised as follows. In Section~\ref{sec:fingerprint-encoding}, we review conventional feature encoding schemes and their limitations in QML tasks of molecular datasets. Section~\ref{sec:encoding-qmse} describes the QMSE approach by defining the hybrid Coulomb–adjacency matrix and its representation as graph states in quantum circuits. Section~\ref{sec:datasets} describes the chemical datasets used for benchmarking, and Section~\ref{sec:results} reports numerical results for classification and regression tasks, highlighting improvements in trainability and generalisation compared to equivalent models using features as inputs. In Section~\ref{sec:discussion}, we discuss the implications of structure encoding in light of recent theoretical findings. Finally, Section~\ref{sec:conclusion} summarises our contributions and outlines future directions, including extensions to quantum kernel methods and expectations for molecular representations in the early quantum fault-tolerant computing regime.

\section{Fingerprint encoding}
\label{sec:fingerprint-encoding}
In supervised QML pipelines, given a dataset $D=\{(\mathbf{x}^1, y^1),\ldots,(\mathbf{x}^M, y^M) \}$ with $M$ pairs of input vectors $\mathbf{x}$ and their corresponding output values $y$, the first step involves the encoding of the input data as initial states in a quantum circuit with $N$ qubits via a chosen feature map. If a variational QML workflow is utilised, as we will employ in this work, a parametrised ansatz circuit with unitary $\hat{U}(\bm{\theta})$ is subsequently applied to the quantum circuit, and the expectation values of a specified Hamiltonian operator $\hat{{H}}$ from each datapoint are obtained to evaluate a given loss function to be minimised classically. The regime will then suggest new parameters to be fed back into the ansatz, and the cycle between quantum and classical interfaces repeats until either a maximum number of iterations  or a suitable convergence criterion is reached \cite{Li2024}. In the context of chemical datasets, where the input molecules in $D$ may be represented in the form of Simplified Molecular Input Line Entry System (SMILES) strings, the chemical moieties and structural features of each input molecule can be further encoded as classical molecular fingerprints, such as RDKit topological fingerprints \cite{Nilakantan1987}.

At first glance it may seem natural to employ basis encoding to map the binary sequences of classical molecular fingerprints, i.e. representing a given data point $\mathbf{x} = (x_1,\ldots,x_{i},\ldots,ex_{\tau})^{\text{T}}$ of fingerprint length $\tau$ as a basis-encoded quantum state, i.e. $\ket{\psi_{BA}}=\ket{x_1\ldots\,x_{i}\ldots\,x_\tau}$, in the computational basis within the Hilbert space $\mathcal{H}_N \cong (\mathbb{C}^2)^{\otimes {N}}$, and applying Pauli-$X$ gates to the corresponding qubits of the fiducial state $\ket{0}^{\otimes N}$ for $x_{i} = 1$ \cite{Balewski2024}. However, in practice, this procedure requires mapping the number of qubits $N$ linearly to the length of each molecular fingerprint,  which  defaults to $\tau=2048$, thus making such a scheme unfeasible. Additionally, although the encoded data points can be expressed efficiently with the lowest possible quantum circuit depth of 1, they only represent a tiny fraction of the total possible number of quantum states within $\mathcal{H}$ with no state overlap. Finding an ansatz that minimises the loss function of the QML task would therefore be an especially challenging endeavour. 

One may instead consider using amplitude encoding to represent the normalised classical molecular fingerprint as a quantum state $\ket{\psi_{AM}}$ in $\mathcal{H}$, thereby drastically reducing the number of qubits required to encode the state to $N=\log_2{\tau}$, or $N=11$ for a default molecular fingerprint. Amplitude encoding also ensures a consistent means of comparing molecules based on the presence or absence of certain chemical moieties, where similar molecular wavefunctions lie in close proximity for $\mathcal{H}$ (and vice versa), thus facilitating the QML task. However, a serious drawback of amplitude encoding is the potential complexity of decomposing the unitary operator $\hat{U}_{AM}$ that evolves the fiduciary state into $\ket{\psi_{AM}}$ in terms of its basis one- and two-qubit quantum gates. Although various quantum gate decomposition schemes have been proposed to prepare any arbitrary quantum state \cite{Mottonen2005, Araujo2021, Shende2006}, such formulations generally require an exponentially increasing number of entangling CNOT gates with $N$, thus making the expression of amplitude-encoded quantum circuits particularly unsuitable for near-term devices.

\begin{figure*}
    \centering
    \includegraphics[width=\linewidth]{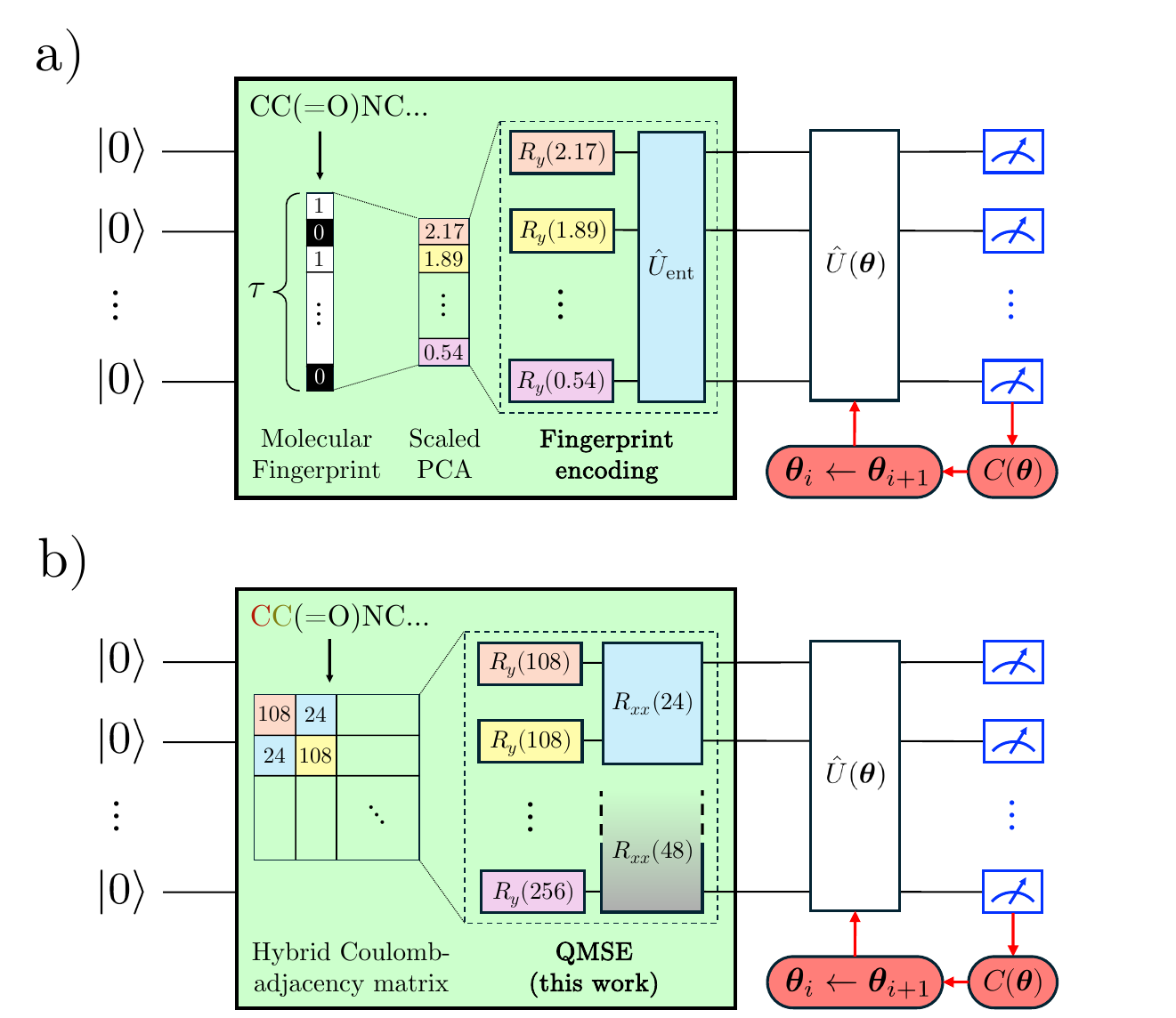}
    \vspace{0.2cm}
    \caption{Schematic of the variational QML workflow for two encoding strategies. a) Fingerprint (angle) encoding: compressed molecular fingerprints are loaded onto the data-encoding layer (in green) as angular rotations. After subsequent evolution of a parameterised ansatz with unitary operator $U(\bm{\theta})$, the circuit's expectation values are evaluated by measuring an observable $\hat{{H}}$, and the resulting cost function $C(\bm{\theta})$ is fed back to a classical optimiser to update the parameters of the ansatz until either the maximum number of iterations or the convergence criterion has been reached. b) Quantum molecular structure encoding: SMILES strings are instead first converted into a hybrid Coulomb‐adjacency matrix and encoded into the quantum circuit by a dedicated data‐encoding layer; the ansatz, measurement of $\hat{{H}}$, and classical parameter‐update loop are then applied.}
    \label{fig:schematic-qmse}
\end{figure*}

To alleviate the drawbacks associated with either basis or amplitude encoding, angle encoding may be employed as an efficient representation of classical data as angular amplitudes for rotational gates in the data-encoding circuit \cite{Salinas2020}. As the default $\tau$ is typically too large to implement angle rotation directly, the number of input features in $\mathbf{x}$ can first be decreased via standard dimensionality techniques, such as principal component analysis (PCA) to its compressed counterpart $\mathbf{\tilde{x}} = ({\tilde{x}}_1 \, \ldots\;  {\tilde{x}}_N)^{\text{T}}$, allowing for a linear scaling in terms of the number of qubits and features. The elements of $\mathbf{\tilde{x}}$ can then be loaded as angles of the rotation gates in the feature map with unitary operator $\hat{U}_{\tilde{\mathbf{x}}}$:
\begin{equation}
\hat{U}_{\tilde{\mathbf{x}}} = \mathlarger{\prod}_{i=1}^{L_\mathbf{x}} \hat{U}_{\rm ent} \Biggl[\mathlarger{\bigotimes}_{j=1}^{N}R_{\hat{P}}({\tilde{x}}_j) \Biggr]
\end{equation}
where $\hat{P} \in \{\hat{X}, \hat{Y}, \hat{Z}\}$ are the Pauli operators, $L_{\mathbf{x}}$ is the number of iterative layers of the data-encoding circuit template, and $\hat{U}_{\rm ent}$ is an optional entangling layer between rotational gates. We will henceforth refer to the process of mapping $\tilde{\mathbf{x}}$ to the data-encoding circuit via angle encoding as \textit{fingerprint encoding} (Fig.~\ref{fig:schematic-qmse}a). Angle encoding has been explored as a flexible means of constructing feature maps in data-encoding circuits for machine learning tasks with real-world datasets on near-term devices, for example, in the ZZFeatureMap scheme \cite{havlivcek2019supervised}. However, as explored in greater detail in Section~\ref{sec:results}, we argue that angle encoding of highly compressed data is generally a poor strategy in QML workflows, largely due to the lack of transferability associated with the mapping of unbounded compressed features into bounded angular amplitudes of rotational quantum gates, thus necessitating alternative encoding schemes for molecular QML tasks that are also qubit- and gate-efficient.

\section{Quantum molecular structure encoding (QMSE)}
\label{sec:encoding-qmse}

\subsection{Hybrid Coulomb-adjacency matrix}
\label{sec:encoding-qmse:matrix}
The Coulomb matrix is commonly used as an intuitive molecular descriptor \cite{neese2003improvement} that encodes the electrostatic interaction between pairs of atoms $(\alpha, \beta)$ within molecules. The diagonal represents a  fit of atomic energies to nuclear charge data, while the off-diagonal elements scale with the interatomic distance $r_{\alpha\beta}$ as $1/r_{\alpha\beta}$. In this work, we use  the \textit{hybrid Coulomb-adjacency} matrix, where we replace $r_{\alpha \beta}$ with the dimensionless parameter $b_{\alpha \beta} \in \{1, 2, 3\}$ depending on the order of the covalent bond. Thus, the modified hybrid Coulomb-adjacency matrix $M_{\alpha \beta}$ is:
\begin{equation}
  M_{\alpha\beta}=\begin{cases}
  \begin{aligned}
    \, 0.5\, {\epsilon_T}\,\mathcal{Z}^{d}_\alpha, \quad & \alpha = \beta \\ \noalign{\vskip7pt}
    \, \frac{{\epsilon_D}\mathcal{Z}_\alpha\mathcal{Z}_\beta}{b_{\alpha\beta}}, \quad & \alpha \neq \beta, \, (\alpha,\beta) \in \mathcal{B} \\ \noalign{\vskip7pt}
    \, 0, \quad \quad & \alpha \neq \beta, \, (\alpha,\beta) \not\in \mathcal{B},
    \end{aligned}
  \end{cases}
\label{eq:main-matrix}
\end{equation}
\noindent where $\mathcal{Z}$ is the atomic number and $b_{\alpha\beta}$ is the bond order defined in the bond set $\mathcal{B}$. The optional parameters ${\epsilon_D}$ and ${\epsilon_T}$ can be specified to differentiate geometric and optical isomers respectively: for the former, ${\epsilon_D}=1$ if a given double bond adopts an \textit{E} configuration and ${\epsilon_D}=-1$ if it adopts a \textit{Z} configuration, while for the latter ${\epsilon_T}=1$ if a given tetrahedral atom is assigned an \textit{R} configuration and ${\epsilon_T}=-1$ if it is assigned an \textit{S} configuration.

The hybrid Coulomb-adjacency matrix differs from the canonical Coulomb matrix in three main aspects, namely:
\begin{itemize}
    \item The off-diagonal elements are non-zero only if atoms $\alpha$ and $\beta$ possess a covalent bond between them in the molecular bond set $\mathcal{B}$, as opposed to the canonical Coulomb encoding, where the off-diagonal matrix elements are generally non-zero due to the long-range interactions of the Coulomb potential, regardless of whether $\alpha$ and $\beta$ are covalently bonded. Therefore, this choice reduces the demands of qubit connectivity when preparing the respective data-encoding quantum circuits.
    
    \item Using $b_{\alpha\beta}$ in place of $r_{\alpha\beta}$ gives rise to off-diagonal elements with larger magnitudes, enabling the data-encoding rotation gates to be much more sensitive and thus facilitating a greater separation between quantum states. Using  $b_{\alpha\beta}$ is also much simpler from an implementation standpoint, whereas $r_{\alpha\beta}$ requires an evaluation of equilibrium bond lengths from standard electronic structure methods to sufficient accuracy.

    \item The exponent of the diagonal elements, $d$, is empirically set to $3.0$ instead of the commonly used value of $2.4$. In our tests, this change was shown to increase the separation of the encoded wave function of molecules (described in further detail in Section~\ref{sec:encoding-qmse:properties}).
\end{itemize}

Thus, we propose hybrid Coulomb-adjacency matrices as a more efficient way to prepare quantum states in QML pipelines.

\subsection{Quantum circuit representation}
\label{sec:encoding-bond:circuit}

In QMSE, molecules within a given chemical dataset are geometrically represented in the data-encoding quantum circuit layer for QML tasks, by mapping the hybrid Coulomb-adjacency matrix as a sequence of one- and two-qubit gates, similar to a QAOA ansatz circuit for connected graphs (Fig.~\ref{fig:schematic-qmse}b). As QMSE depends heavily on the atomic identities and covalent connectivity of each molecule, it is expected to produce more distinct representations of the molecular structure compared to fingerprint encoding. The unitary operator that describes the QMSE quantum circuit for a given molecule is:
\begin{equation}
\hat{U}_{\mathbf{x}} = \mathlarger{\prod}_{k=1}^{L_{\mathbf{x}}} \biggl\{ 
\mathlarger{\bigotimes}^{|\mathcal{B}|}_{N_\alpha<N_\beta} R_{\hat{P}^2}\bigl(M_{\alpha\beta}\bigr) \,\mathlarger{\bigotimes}_{i=1}^{N_{\alpha}} R_{\hat{P}}\bigl(M_{ii}\bigr) \biggr\}
\label{eq:unitary}
\end{equation}
\noindent where $R_{\hat{P}} \in \{R_x,\, R_y,\, R_z \}$ and $R_{\hat{P}^2} \in \{R_{xx},\, R_{yy},\, R_{zz} \}$ are the one- and two-qubit parameterised rotation gates representing the atoms and bonds of the molecule, respectively. {It should be noted that $N$ should be greater than or equal to the length of the encoded SMILES string. We now prove the following:}
{\begin{theorem}
There exists a bijection $\phi:\mathcal{S}\rightarrow\mathcal{Q}$ between the set of SMILES strings $\mathcal{S}$ and the set of QMSE unitaries $\mathcal{Q}$ for a given $L_\mathbf{x}$ and $N$.
\end{theorem}}
{\begin{proof}
It is easy to see that $\phi$ is surjective, since for every constructed QMSE unitary operator there exists at least one corresponding SMILES string. For injectivity, suppose $\phi(s_1) = \phi(s_2)$ where $s_1, s_2 \in \mathcal{S}$, or equivalently $\hat{U}_{s_1} = \hat{U}_{s_2}$, where $\hat{U}_{s_1}, \hat{U}_{s_2} \in \mathcal{Q}$. This result implies that the two QMSE unitaries representing $s_1$ and $s_2$ are composed of the same $R_{\hat{P}}$ and $R_{\hat{P^2}}$ with the same angular values. As each encoded rotational angle $\tilde{x}$ is not periodic, i.e. $\tilde{x}_1 \neq \tilde{x}_2 - a\pi$ for some $a\in \mathbb{Z}$, all encoded angles are uniquely interspersed between $[-2\pi, 2\pi)$. Hence, each one- and two-qubit rotation gate uniquely represent each atom and bond of a given molecule respectively. Moreover, since the corresponding $R_{\hat{P^2}}$ operators commute with one another, the ordering of the two-qubit gates does not matter even if a different permutation is chosen to populate the two-qubit gates. Therefore, $\hat{U}_{s_1}$ and $\hat{U}_{s_2}$ must necessarily encode the same SMILES string, i.e. $s_1 = s_2$. Thus, $\phi$ is bijective.
\end{proof}}

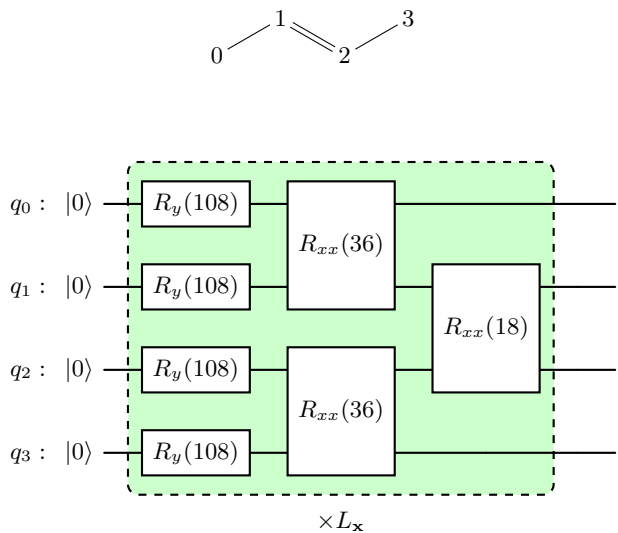
\begin{figure}
    \centering
    \setchemfig{angle increment=30}
    \chemname{
        \chemfig{0-[1]1=[-1]2-[1]3}
    }{}
    \vspace{30pt}
    
    \begin{quantikz}
        \lstick{$q_0:~\ket{0}$}  & \gate{R_y(108)}\gategroup[4,steps=3,style={dashed,rounded
        corners,fill=green!20, inner
        xsep=2pt},background,label style={label
        position=below,anchor=north,yshift=-0.3cm}]{{ $\times L_{\mathbf{x}}$ }} &  \gate[2]{R_{xx}(36)} & & & \\
        \lstick{$q_1:~\ket{0}$} & \gate{R_y(108)} &  & \gate[2]{R_{xx}(18)} & & \\
        \lstick{$q_2:~\ket{0}$} & \gate{R_y(108)} & \gate[2]{R_{xx}(36)} & & & \\
        \lstick{$q_3:~\ket{0}$} & \gate{{R_y}(108)} &  & & &
    \end{quantikz}
    \vspace{10pt}
    
    \caption{Example molecular encoding layer (bottom) forming part of a 4-qubit quantum circuit representing an (\textit{E})-but-2-ene molecule (top) using $R_y$ and $R_{xx}$ gates with $L_{\mathbf{x}}$ number of data-encoding layers. The rotational angles associated with the gates are tuned based on the elements of the hybrid Coulomb-adjacency matrix in Eq.~\ref{eq:main-matrix}.}
    \label{fig:2butene}
\end{figure}

Furthermore, the choice of $\hat{P}^2$ in the construction of the two-qubit quantum gates is also useful as it allows for rearrangement during transpilation to produce the same wavefunction with a lower circuit depth. This effect can be illustrated in Fig.~\ref{fig:2butene} with a four-qubit QMSE circuit for the molecule (\textit{E})-but-2-ene with a SMILES string representation of CC/=/CC. {From Eq. 2, the hybrid Coulomb-adjacency matrix with a qubit ordering of $\ket{q_0q_1q_2q_3}$ of (\textit{E})-but-2-ene can be written as:
\begin{equation}
M_{\alpha\beta} = \begin{bmatrix}
\;108 & 36 & 0 & 0\;\; \\
\;36 & 108 & 18 & 0\;\; \\
\;0 & 18 & 108 & 36\;\; \\ 
\;0 & 0 & 36 & 108\;\;
\end{bmatrix}
\end{equation}}

{Designating $R_y$ and $R_{xx}$ as the encoded one- and two-qubit quantum gates respectively}, the encoded $R_{xx}$ gate between qubits $q_2$ and $q_3$ can be transpiled such that it can be run simultaneously with the $R_{xx}$ gate between $q_0$ and $q_1$ without changing $\hat{U}_{\mathbf{x}}$, due to the commutativity of the $R_{xx}$ gates.

\subsection{Properties of QMSE quantum circuits}
\label{sec:encoding-qmse:properties}
{From the previous theorem, as a bijection exists between the unitary operators of QMSE and SMILES strings, QMSE unitaries have similar properties with those of SMILES strings, and by extension, classical Coulomb and hybrid-Coulomb adjacency matrices.} Notably, QMSE operators are invariant to molecular translations and rotations in $SO(3)$; however, they are not permutationally invariant, as reordering the atom-mapped qubits results in different unitary matrices \cite{Montavon2012}. 

Another similarity between Coulomb matrices and QMSE operators is the treatment of unrepresented atoms in smaller molecules. Coulomb matrix entries are filled with  zeros up to the required maximum number of atoms, while for QMSE, a virtual identity gate is instead implemented for each unrepresented qubit. However, a major difference lies in computational complexity. The matrix entries associated with unrepresented atoms of Coulomb matrices are typically used as input elements in the classical ML regime, while the computational cost of implementing unrepresented qubits in QMSE circuits is essentially zero, as the qubits have already been set to the fiduciary state. This difference produces a highly efficient linear scaling of the combined number of atoms and bonds in a given molecule with the number of data-encoding quantum gates, allowing less complex molecules in a given dataset to be represented by simpler QMSE operators, and vice versa.

A basic problem in cheminformatics is the comparison of chemical moieties between two molecules, which we will label as $P$ and $Q$ for illustration. The comparison can be carried out classically by calculating the chemical similarity. One of the most popular methods involves computing the Tanimoto similarity \cite{Bajusz2015} between the corresponding molecular fingerprints of $P$ and $Q$, $\mathbf{x}_P$ and $\mathbf{x}_Q$, respectively:
\begin{equation}
T(P,Q)=\frac{|\mathbf{x}_P \cup \mathbf{x}_Q|}{|\mathbf{x}_P|+|\mathbf{x}_Q|-|\mathbf{x}_P \cap \mathbf{x}_Q|}.
\end{equation}
In the quantum picture, we instead quantify the chemical similarity between $P$ and $Q$  via the quantum overlap or fidelity $F$ between the corresponding wavefunctions $\ket{\psi_P}$ and $\ket{\psi_Q}$:
\begin{align}
\begin{split}
    F(P,Q)&=|\braket{\psi_Q}{\psi_P}|^2 \\
    &= |\bra{\mathbf{0}}\hat{U}^{\dagger}_Q\hat{U}_P\ket{\mathbf{0}}|^2 .
\end{split}
\end{align}
We show that the following property holds for the fidelity of extended molecular chains:

\begin{property}
    Let the SMILES string representations of $P$ and $Q$ be ordered as $p_1\alpha-\beta{p_2}$ and $q_1\alpha-\beta{q_2}$, with common atoms $\alpha$ and $\beta$ bonded with the same bond order and mapped to the same qubits. Now consider the extended molecular representations of $P$ and $Q$, $\tilde{P}$ and $\tilde{Q}$ respectively, with some common molecular fragment $\mathbf{c}$ mapped to the same qubit arrangement, i.e. $\tilde{P} = p_1\alpha-\mathbf{c}-\beta{p_2}$ and $\tilde{Q} = q_1\alpha-\mathbf{c}-\beta{q_2}$. Then for $L_{\mathbf{x}}=1$, $F(\tilde{P}, \tilde{Q})=F(P,Q)$.
\end{property}

\begin{figure}
    \centering
    \includegraphics[width=\linewidth]{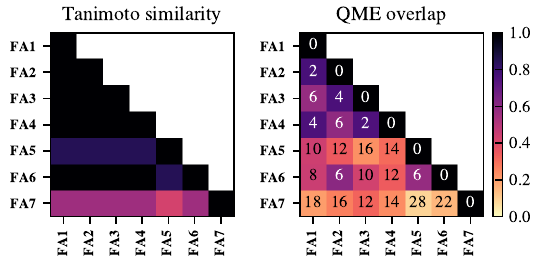}
    \caption{Tanimoto similarity (left) and QMSE fidelity (right) of the fatty acids series \textbf{FA1$-$FA7}. The chemical overlap is computed from SMILES strings encoded via QMSE, using the default $R_y + R_{xx}$ combination. For each overlap pair, the number of qubits of the unitary circuits is also reported after chain contraction.}
    \label{fig:overlap:fattyacids}
\end{figure}

\begin{proof}
Let $\hat{V}$ and $\hat{W}$ be the unitary QMSE representations of the one- and two-qubit rotation layers, respectively. Define:
\begin{gather*}
\hat{V}_P=\hat{V}_{p_1}\otimes\hat{V}_{\alpha}\otimes\hat{V}_{\beta}\otimes\hat{V}_{p_2} \\
\hat{V}_Q=\hat{V}_{q_1}\otimes\hat{V}_{\alpha}\otimes\hat{V}_{\beta}\otimes\hat{V}_{q_2} \\
\hat{V}_{\tilde{P}}=\hat{V}_{p_1}\otimes\hat{V}_{\alpha}\otimes\mathbbm{1}
_{\mathbf{c}}\otimes\hat{V}_{\beta}\otimes\hat{V}_{p_2} \\
\hat{V}_{\tilde{Q}}=\hat{V}_{q_1}\otimes\hat{V}_{\alpha}\otimes\mathbbm{1}_{\mathbf{c}}\otimes\hat{V}_{\beta}\otimes\hat{V}_{q_2} \\
\hat{U}_c = \hat{W}_{\mathbf{c}\backslash\{\alpha, \beta\}}\hat{V}_c
\end{gather*}
It  follows that 
\begin{align*}
\begin{split}
F&(\tilde{P},\tilde{Q}) =\\
&= |\bra{\tilde{\mathbf{0}}}\hat{U}^{\dagger}_{\tilde{Q}}\hat{U}_{\tilde{P}}\ket{\tilde{\mathbf{0}}}|^2\\
&= 
|\bra{\tilde{\mathbf{0}}}\hat{V}^{\dagger}_{\tilde{Q}}\hat{W}^{\dagger}_{\tilde{Q}\backslash \mathbf{c}}\hat{U}^{\dagger}_{\mathbf{c}}\hat{W}^{\dagger}_{\alpha{\mathbf{c}}}\hat{W}^{\dagger}_{\beta{\mathbf{c}}}\hat{W}_{\beta{\mathbf{c}}}\hat{W}_{\alpha{\mathbf{c}}}\hat{U}_{\mathbf{c}}\hat{W}_{\tilde{P}\backslash \mathbf{c}} \hat{V}_{\tilde{P}}
 \ket{\tilde{\mathbf{0}}}|^2 \\
 &= | \bra{\tilde{\mathbf{0}}} 
\hat{V}^{\dagger}_{\tilde{Q}}\hat{W}^{\dagger}_{\tilde{Q}} \hat{W}_{\tilde{P}} \hat{V}_{\tilde{P}}
 \ket{\tilde{\mathbf{0}}}|^2 \\
  &= | \bra{{\mathbf{0}}} 
\hat{V}^{\dagger}_{{Q}}\hat{W}^{\dagger}_{{Q}} \hat{W}_{{P}} \hat{V}_{{P}}
 \ket{{\mathbf{0}}}|^2 \\
 &=F(P,Q).
\end{split}
\end{align*}

\end{proof}
This property is instrumental in setting up the following corollary for computing the fidelity between molecules with common substructures using a reduced number of qubits:
\begin{corollary}
If two given molecular representations $\tilde{P}$ and $\tilde{Q}$ share some common molecular fragments $\mathcal{C}=(\mathbf{c}_1, \mathbf{c}_2,...)$ bonded identically to the same common atoms mapped to the same qubits, then $F(\tilde{P}, \tilde{Q})$ with ${N}$ qubits can be  evaluated with  $N-{N_{\mathcal{C}}}$ qubits via $F({P}, {Q})$ by eliminating $\mathcal{C}$ in both $\tilde{P}$ and $\tilde{Q}$.
\end{corollary}
We will refer to the process set out in the corollary as \textit{chain contraction}. {This procedure is guaranteed for QMSE circuits with $L_x=1$, regardless of the choice of $R_{\hat{P}}$ and $R_{\hat{P}^2}$.} To highlight the effectiveness of chain contraction, we numerically computed the quantum fidelities of a curated series of seven unsaturated fatty acids labelled {\textbf{FA1$-$FA7}}, each containing 34 carbon atoms (Fig.~\ref{fig:overlap:fattyacids}), and compared the result to the  classical Tanimoto similarity (Fig.~\ref{fig:overlap:fattyacids}). Although the Tanimoto  metric was generally able to distinguish {\textbf{FA7}}, with  a larger degree of unsaturation, from the rest of the  structures, the long saturated chains of 
\textbf{FA1$-$FA6}
 make them difficult to distinguish.
 Quantum fidelities are  more effective in evaluating chemical similarity, reflecting the  wide variability of unsaturation and the double bond positions. Although the evaluation of quantum fidelities would ordinarily be computationally intensive due to the large number of atoms per molecule,  eliminating common molecular fragments via chain contraction allows for a potentially large reduction in the number of implemented qubits and quantum gates. Fig.~\ref{fig:overlap:fattyacids} illustrates the number of qubits required to calculate each fidelity pair. The structures of {\textbf{FA1$-$FA7}} and the chain contraction procedure are outlined in Appendix~\ref{app:fatty-acids}.

\section{Datasets}
\label{sec:datasets}
We compiled a dataset of 105 linear saturated small organic molecules from the CRC Handbook of Chemistry and Physics (95\textsuperscript{th} Edition) \cite{haynes2014crc}. This dataset includes 50 alkanes, 38 monohydric alcohols, and 17 monohydric ethers with varying degrees of positional isomerism. Canonical SMILES for each molecule were first constructed via the RDKit canonicalisation algorithm. The alcohol and ether SMILES strings were then reordered with the oxygen atom in the left-most position, to maximise fidelities between similar chemical moieties and vice versa. To assess the performance of the algorithm with datasets of increasing size and complexity, we further partitioned the complete dataset of 105 molecules into two subsets: the \textit{alkane} subdataset with only the alkane structures, and the \textit{oxygen} subdataset with only the alcohol and ether structures.

To better understand chemical similarity within the QMSE framework, we systematically benchmarked the effect of different combinations of one- and two-qubit rotation gates on the fidelities of the chemical datasets, as defined in Eq. 5. We fixed the single‐qubit rotation to $R_y$ gates and varied the two‐qubit entangling operations among $R_{xx}$, $R_{yy}$, and $R_{zz}$. All simulations were performed with $L_{\mathbf{x}} = 1$ under noiseless statevector conditions for different pairs of chain-contracted molecules within the alkane- and oxygen-subdatasets. 

\begin{figure*}
    \centering
    \includegraphics[width=\linewidth]{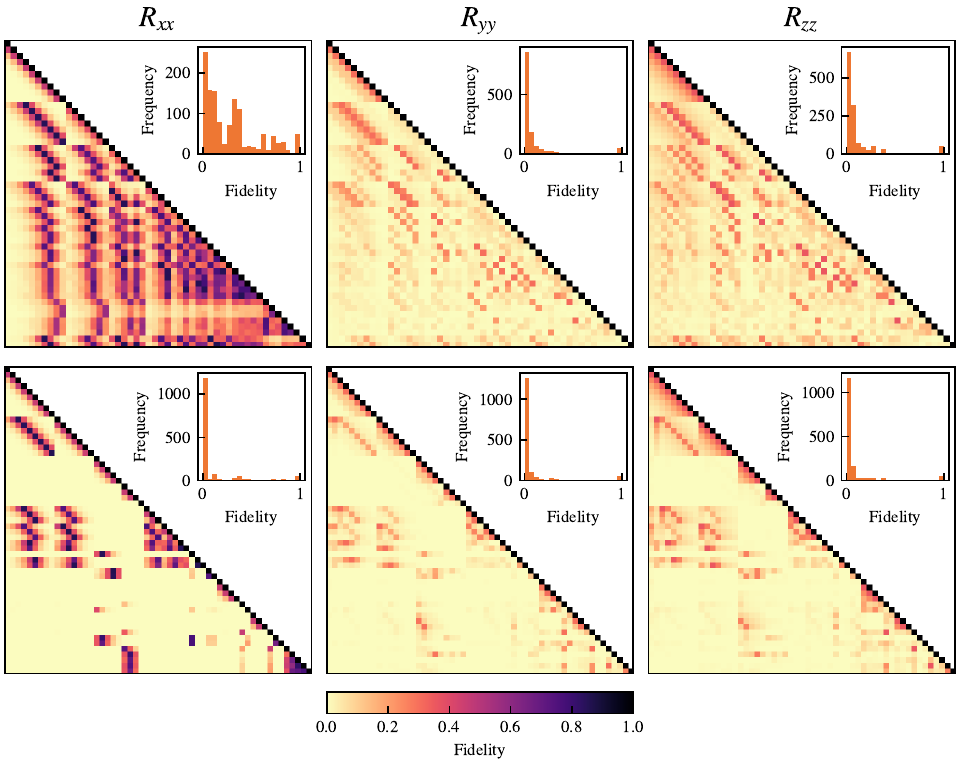}
    \caption{Heatmaps of fidelity matrices for chain-contracted molecules encoded with QMSE within the alkane- (top row) and oxygen- (bottom row) subdatasets. The one-qubit data-encoding gate for all configurations is fixed as $R_y$, while the two-qubit data-encoding gate is varied to produce the fidelity matrix for $R_{xx}$ (left column), $R_{yy}$ (middle column), and $R_{zz}$ (right column). The colour corresponds to the fidelity of each molecule pair within their respective insets. Based on the overall distribution of the fidelity values, $R_{y}+ R_{xx}$ is selected as the default setup for QMSE.}
    \label{fig:overlap}
\end{figure*}

Fig.~\ref{fig:overlap} shows the heatmaps of quantum fidelity matrices for the alkane- and oxygen-subdatasets. In both cases, the $R_y + R_{xx}$ configuration displays the widest range of fidelities,  compared to   $R_y + R_{yy}$ or $R_y + R_{zz}$. This separation directly translates into improved discrimination capability in QML tasks. Accordingly, we use $R_y$ for single‐qubit rotations and $R_{xx}$ for two‐qubit entangling gates as our default encoding setup for all subsequent experiments using QMSE.

\section{Results}
\label{sec:results}

\begin{table*}
   \caption{Summary of classification and regression runs with the variational approach of Section~\ref{sec:results}. 
   For the data-encoding layer, the configuration used for fingerprint encoding employs an initial $R_y$ rotation gate layer followed by a linear chain of CNOT gates with $L_{\mathbf{x}} = 1$; for QMSE the configuration utilises the default $R_y + R_{xx}$ configuration with $L_{\mathbf{x}} = 1$.
   From left to right, the columns indicate the run ID, type of task assigned to the machine learning model (classification or regression), molecular dataset, ansatz configuration, Hamiltonian, and the maximum number of COBYLA iterations.} 
   \label{tab:mk-runs}
   \small
   \centering
   \begin{tabular}{clllcccccc}
   \toprule
    \textbf{Run ID} &\textbf{Task} & \textbf{Dataset} & \textbf{Encoding} & \multicolumn{4}{c}{\textbf{Ansatz configuration}} & Hamiltonian & \textbf{Iterations} \\
     &&&& 1-qubit & 2-qubit & Entanglement & Layers & & $(~/ 10^3~)$  \\ 
   \midrule
   1 & Classification & Alkane & Fingerprint & $R_y$ & CZ & Linear & $[1-5]$ & Global all-$\hat{Z}$& 1 \\
   2 & Classification & Alkane & Fingerprint & $R_y$ & CZ & Pairwise & $[1-5]$ & Global all-$\hat{Z}$& 1 \\
   3 & Classification & Alkane & QMSE & $R_y$ & CZ & Linear & $[1-5]$ & Global all-$\hat{Z}$& 1 \\
   4 & Classification & Alkane & QMSE & $R_y$ & CZ & Pairwise & $[1-5]$ & Global all-$\hat{Z}$& 1 \\
   \midrule
   5 & Classification & Complete & QMSE & $R_y$ & CZ & Pairwise & $[1-5]$ & Global all-$\hat{Z}$& 2 \\
    6 & Classification & Complete & QMSE & $R_y$ & CRX & Pairwise & $[1-5]$ & Global all-$\hat{Z}$& 2 \\
    7 & Classification & Complete & QMSE & $R_y$ & CRX & Pairwise & $[1-5]$ & $IIIIZZIIII$ & 2 \\ 
    8 & Classification & Complete & QMSE & $R_y$ & CRX & Pairwise & $[1-5]$ & $IIIZZZZIII$ & 2 \\ 
    9 & Classification & Complete & QMSE & $R_y$ & CRX & Pairwise & $[1-5]$ & $ZZIIIIIIII$ & 2 \\
    \midrule
   10 & Regression & Alkane & QMSE & $R_y$ & CRX & Pairwise & $[1-6]$ & Global all-$\hat{Z}$ & 10 \\
   11 & Regression & Alkane & QMSE & $R_y$ & CRX & Full & $[1-6]$ & Global all-$\hat{Z}$ & 10 \\
   \bottomrule
   \end{tabular}
\end{table*}

We perform three main QML numerical experiments to investigate the effectiveness of QMSE,  summarised in Table~\ref{tab:mk-runs}. The first task involves a binary classification of the alkane subdataset, where we predict whether a given molecule is in the gas phase at 100~$^{\circ}$\,C. We contrast the default $R_y+R_{xx}$ QMSE configuration (Runs 3-4) with standard fingerprint encoding (Runs 1-2). For the latter runs, the data was preprocessed by converting the subdataset SMILES strings into RDKit topological molecular fingerprints comprising 2048 bits each, and subsequently reducing each fingerprint via PCA into ten coordinates, corresponding to the same maximum qubit size of the data-encoding layer of QMSE. The fingerprint encoding layer was then prepared with a single layer of $R_y$ gates and rotation angles corresponding to the PCA-reduced coordinates scaled in the range $[-2\pi, 2\pi]$, followed by a linear entangling layer of CNOT gates. Both the fingerprint-encoded and QMSE data-encoding layers are  followed by a variational ansatz composed of $R_y$ gates followed by an entangling layer $\hat{U}_{\text{ent}}$, where $\hat{U}_{\text{ent}}$ is either a linear arrangement of CZ gates (Runs 1, 3) or a pairwise arrangement of CZ gates (Runs 2, 4):
\begin{equation}
\hat{U}(\bm{\theta}) = \mathlarger{\prod}_{i=1}^{L_\mathbf{\theta}} \hat{U}_{\rm ent} \Biggl[\mathlarger{\bigotimes}_{j=1}^{N}R_{y}({{\theta}}_j) \Biggr],
\end{equation}
where $\hat{U}_{\rm ent}$ can be configured to express an ansatz with full, linear, and pairwise entanglement, and $L_{\theta}$ is the number of ansatz layers.

The variational quantum classifier (VQC) circuit is then measured in the global all-$\hat{Z}$ basis and optimised via a gradient-free COBYLA regime for a maximum  of 1,000 iterations with an $L_2$ loss function,  assigning $y=1$ for predicted $\langle \hat{H} \rangle$ values greater than 0 and $y=0$ for predicted $\langle \hat{H} \rangle$ values less than 0.

To further evaluate the performance of QMSE for a wider range of chemical moieties, we broaden the same classification task to the complete dataset, including monohydric alcohol and ether molecules (Runs 5-9), and increase the maximum number of COBYLA iterations to 2,000 for improved convergence. As the classification task complexity is increased, we compare the effect of the variational ansatz entangling CZ gate (Run 5) with a more expressive controlled-$R_x$ (CRX) gate (Run 6) with pairwise entangling configurations. We also vary the Hamiltonian measured by the VQC circuit by experimenting with local Hamiltonians, where we measure a selection of qubits in the $\hat{Z}$-basis (Run 7-9), as local cost functions have been shown to be more trainable than global cost functions in parameterised quantum circuits up to $L_{\theta} \in \mathcal{O}(\log N)$ \cite{Cerezo2021}. In particular, we wish to explore the light-cone phenomenon arising from the pairwise entangling arrangement of the ansatz \cite{Abedi2023}, whose effect is more pronounced when measuring the middle two qubits (Run 7) and the middle four qubits (Run 8) in the default $\hat{Z}$-basis. As the oxygen subdataset contains alcohols and ethers that predominantly feature the oxygen atom in the first two qubit positions, we also seek to understand the impact of measuring the first two qubits in the $\hat{Z}$-basis (Run 9) compared to the other local Hamiltonian runs.  

Finally, we tackle the much more difficult task of regression using the alkane subdataset to predict  boiling points via a variational quantum regressor (VQR). Here, we normalise the boiling points (in Kelvin) to  the range $[-0.5, 0.5]$, rather than $[-1, 1]$, as the difficulty for the QML model to express expectation values with larger magnitudes is much higher. Moderating the range of predicted expectation values reduces the risk of underfitting molecules with the lowest or highest boiling points. This choice also allows the model to potentially extrapolate boiling points outside the predicted range. We benchmark the performance with an $R_y+\text{CRX}$ variational ansatz, with either a pairwise (Run 10) or a full entanglement (Run 11) configuration. The VQR circuit was then measured in the global all-$\hat{Z}$ basis and optimised for a maximum number of 10,000 COBYLA iterations with an $L_2$ loss function.

All runs were evaluated via stratified $k$-fold cross validation, with the alkane and complete datasets split into $k=5$ equally sized groups of samples. The classification tasks were assessed with the mean accuracy scores of both training and test datasets, while the regression tasks were evaluated with the coefficient of determination $R^2$ of both training and test datasets. For each cross-validated iteration, a total of 100 random initial coordinates $\bm{\theta}_i\in[-2\bm{\pi}, 2\bm{\pi}]$ for the variational ansatz were sampled, and the median results were tabulated across the number of initial coordinates and the number of $k$-fold samples. This process was  repeated across a discrete range of ansatz layers for each run, with the classification tasks ranging between 1-5 ansatz layers and the regression task between 1 and 6 ansatz layers.

\subsection{Classification}
\label{sec:qml:classification}

\begin{figure*}[t]
    \centering
    \includegraphics[width=\textwidth]{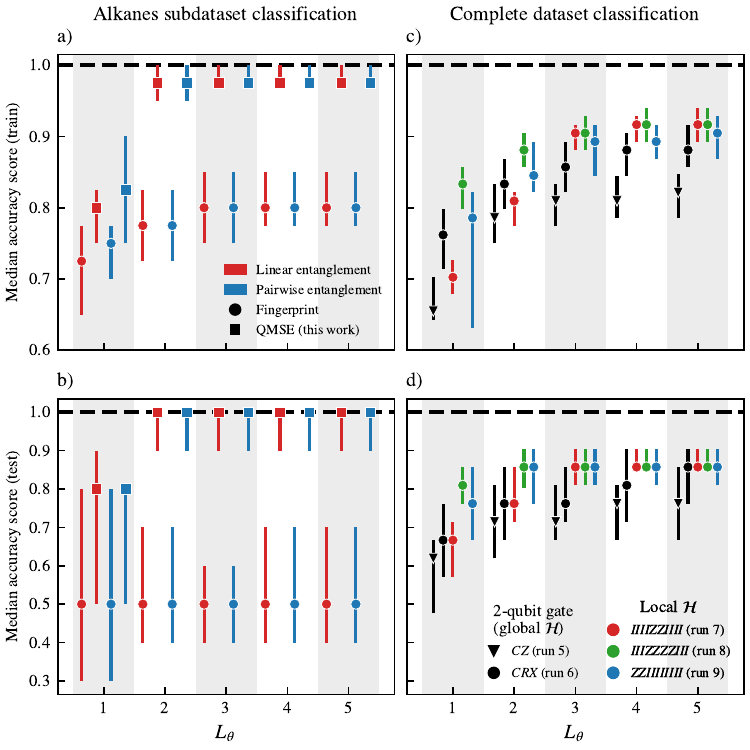}
    \caption{Accuracy scores when classifying molecules in the alkane (left, a and b) and complete (right, c and d) datasets for Runs 1$-$9 with 1$-$5 $L_{\theta}$ number of ansatz layers. The top row shows the median training accuracy scores, and the bottom row shows the median test accuracy scores. The error bars indicate the 16\textsuperscript{th} and 84\textsuperscript{th} percentile values of the average accuracies of the $k$-fold splits.}
    \label{fig:classification-scores}
\end{figure*}

For the classification task on the alkane subdataset, QMSE achieves excellent results with a consistently high training and corresponding test accuracy score for increasing $L_{\theta}$ (Fig.~\ref{fig:classification-scores}a, ~\ref{fig:classification-scores}b). At $L_{\theta}=3$ and above, aside from the data outliers of 2,2,3,3-tetramethylpentane and occasionally methane, the VQC model coupled with QMSE was trained with perfect accuracy scores, and generalised to  unseen data in the test splits with minimal overfitting. In contrast,  the modest improvement in training accuracy scores for the VQC model coupled fingerprint encoding with increasing $L_{\theta}$ translates poorly to the test splits with low accuracy scores. This result strongly indicates that fingerprint encoding coupled to  PCA-reduced data is a poor data encoding framework for representing molecular structures in chemical QML tasks. The stark contrast in performance between fingerprint encoding and QMSE can also be attributed to the difference in their respective loss curves, where QMSE converges to a much lower loss (refer to Appendix ~\ref{app:loss-functions} for more details on the loss curves).
For different  ansatz entanglement schemes with both fingerprint encoding and QMSE, there appears to be no significant difference between linear and pairwise entanglement in terms of accuracy scores. 

Extending our exploration to the classification of the complete dataset, we observe good performance in the training of the VQC model with QMSE and  generalisation to the test samples (Fig.~\ref{fig:classification-scores}c, ~\ref{fig:classification-scores}d). Owing to the added complexity of the expanded dataset with alcohol and ether moieties, we found that modifying the ansatz entangling gate from CZ (Run 5) to CRX (Run 6) produced a marked increase in training and test accuracy results. Modification of the Hamiltonian from a global all $\hat{Z}$-basis to a local $\hat{Z}$-basis further improved performance at higher $L_{\theta}$. This result is particularly impressive, especially from a quantum error perspective, as reducing the number of measured qubits also reduces the source of noise attributed to crosstalk between qubits \cite{Rodriguez2021}. In particular, Runs 7-8 attain slightly higher training accuracies of over 90\% at $L_{\theta} = 5$  compared to Run 9, suggesting that the light-cone phenomenon is a little more significant in training VQCs than the emphasis in measuring the first two qubits.

\subsection{Regression}
\label{sec:qml:regression}

\begin{figure*}[t]
    \centering
    \includegraphics[width=\textwidth]{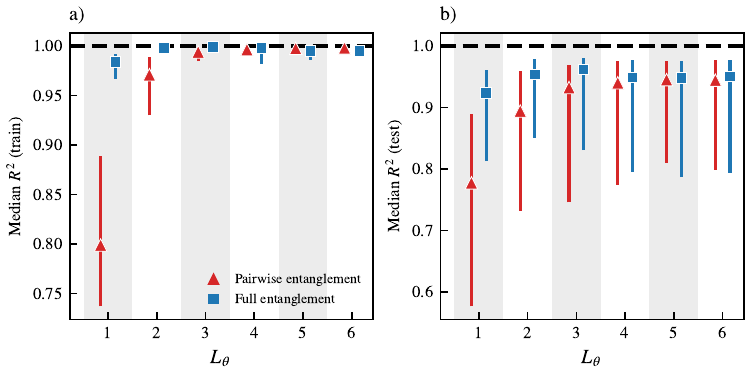}
    \caption{$R^2$ scores for boiling point prediction of the alkane subdataset for Runs 10$-$11 with 1$-$6 $L_{\theta}$ number of ansatz layers, with a) depicting the median training scores (left), and b) the median test scores (right). The error bars depict the 16\textsuperscript{th} and 84\textsuperscript{th} percentile values of the average $R^2$ values of the $k$-fold splits.}
    \label{fig:regression-scores}
\end{figure*}

Lastly, by considering the alkane regression task (Fig. ~\ref{fig:regression-scores}), we find excellent training $R^2$ scores, however the generalisation of the trained VQR model to the test data reaches a limit with increasing $L_{\theta}$. This result suggests a proliferation of local minima with lower $R^2$ scores arising from different starting ansatz coordinates having a larger effect on the more difficult task of predicting continuous variables. Nevertheless, this is still an encouraging result with $R^2$ test values exceeding 0.95 for optimal starting conditions with little to no overfitting. In terms of ansatz arrangement, Run 11 with full entanglement appears to perform slightly better in terms of both $R^2$ training and test scores compared to Run 10 with pairwise entanglement, which is unsurprising given the increase in parameters for the same $L_{\theta}$.

\section{Discussion}
\label{sec:discussion}

Our results show that QMSE greatly outperforms conventional fingerprint encoding in the alkane classification task, in addition to performing quite well for more complex classification and regression tasks. The improved gains observed with QMSE arise, not only from improved state separability, but also from its inherent interpretable structure, mirroring key advantages that are also desirable in classical machine learning. In classical ML, interpretability enables understanding of how input features affect model decisions, guiding both model debugging and feature engineering. For QML, where feature maps are implemented as quantum circuits, the data-encoding scheme governs the entire hypothesis space, and thus understanding its structure is especially critical in designing more expressive and highly trainable models.

\begin{figure}[h!]
    \centering
    \includegraphics[width=\linewidth]{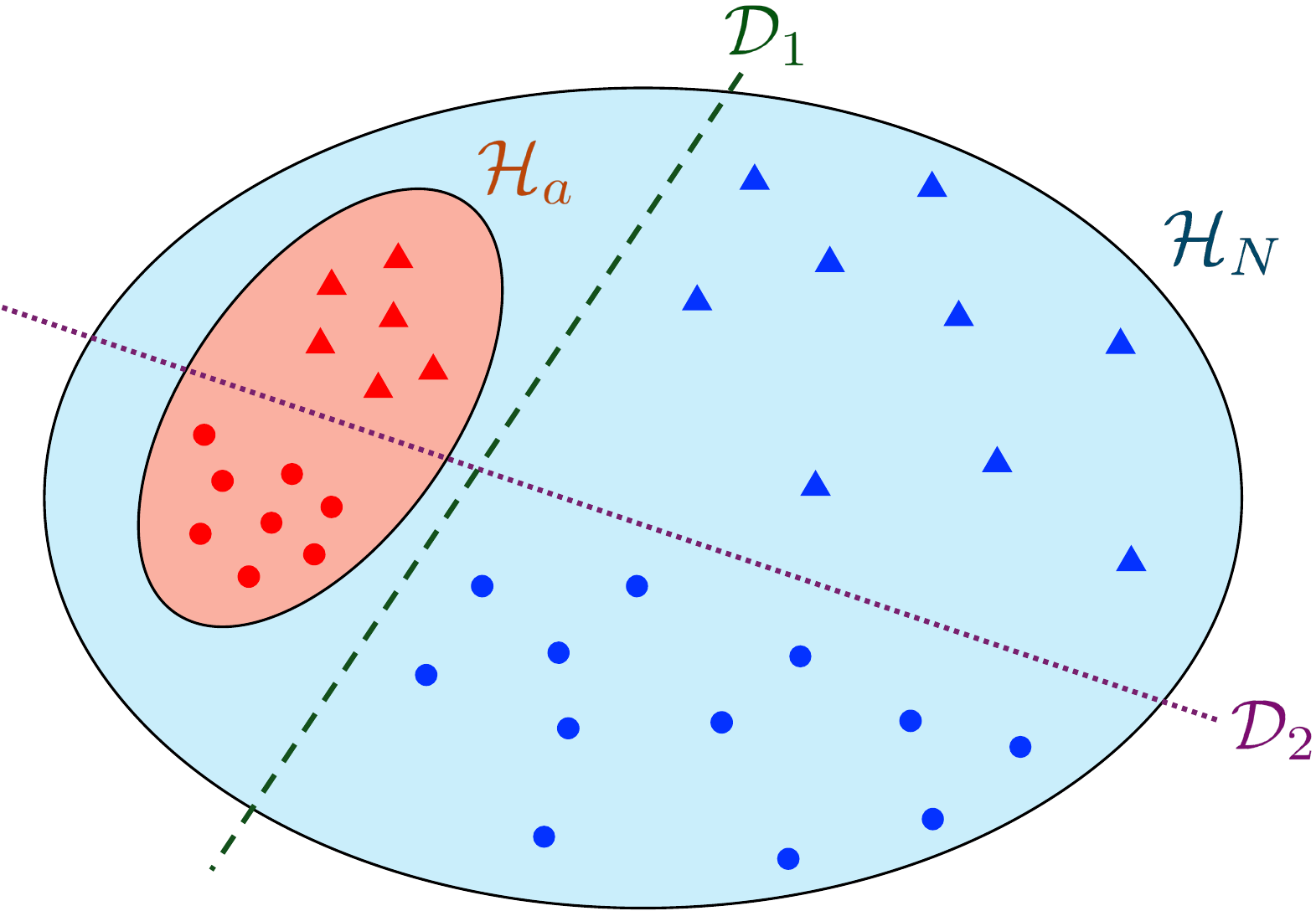}
    \caption{Illustration of decision boundaries $\mathcal{D}_1$ and $\mathcal{D}_2$ separating a molecular datatset encoded with QMSE in $\mathcal{H}_N$. $\mathcal{D}_1$ (green line) aims to separate the datapoints encoding smaller molecules in $\mathcal{H}_a$ (red) from the larger molecules outside $\mathcal{H}_a$ (blue), while $\mathcal{D}_2$ (purple line) seeks to separate datapoints based on atomic or bond identities (circles vs. triangles).}
    \label{fig:discussion}
\end{figure}

The interpretability of QMSE can be rationalised by considering two main decision boundaries maximising separation between datapoints with different properties (Fig.~\ref{fig:discussion}). Consider a molecular dataset encoded with QMSE, with their resulting statevectors residing in the Hilbert space $\mathcal{H}_N$. Now consider the Hilbert space $\mathcal{H}_a = \mathcal{H}_{N-1}\otimes{\mathcal{H}_{\ket{0}}}$, such that $\mathcal{H}_a\subset\mathcal{H}_N$. The molecular statevectors in $\mathcal{H}_a$ with at least one less encoded atom than the maximum number of encoded qubits can thus be separated from the maximally encoded statevectors lying outside $\mathcal{H}_a$ with the decision boundary $\mathcal{D}_1$, and this reasoning extends for increasingly small subspaces of $\mathcal{H}_a$ that smaller molecules reside in. To further distinguish between atomic and bond identities, a second decision boundary $\mathcal{D}_2$ can be used that cuts across $\mathcal{H}_a$ and $\mathcal{H}_{N} \cup \mathcal{H}_{a}'$. For example, consider some molecular dataset $S = \{\ket{\psi_{\text{CC}}}, {\ket{\psi_{\text{CCC}}}}, \ket{\psi_{\text{C=C}}}, \ket{\psi_{\text{OC}}}, {\ket{\psi_{\text{C=CC}}}}, \ket{\psi_{\text{OCC}}}\}$ in $\mathcal{H}_N$ for $N=3$.  $\mathcal{D}_1$ can be established between molecular subsets with two atoms and three atoms, i.e. $S_1=\{\ket{\psi_{\text{CC}}}, \ket{\psi_{\text{C=C}}}, \ket{\psi_{\text{OC}}}\}$ and $S_2=\{\ket{\psi_{\text{CCC}}}, \ket{\psi_{\text{C=CC}}}, \ket{\psi_{\text{OCC}}}\}$, where $S_1 \subset \mathcal{H}_{a}$ and $S_2 \subset \mathcal{H}_{N} \cup \mathcal{H}_{a}'$. For $\mathcal{D}_2$, the datapoints can be further partitioned based on the bond order of the first C$-$C bond, as well as the identity of the first atom.

Quantum state overlaps of different molecules derived from QMSE exhibit higher variance and wider spread compared to those from fingerprint encodings (Fig.~\ref{fig:overlap:fattyacids}-\ref{fig:overlap}), reflecting better state distinguishability and fewer kernel concentration issues \cite{thanasilp2024exponential}. This result suggests that QMSE defines a more meaningful quantum feature space, consistent with theoretical prescriptions for quantum advantage \cite{liu2021rigorous}.

Finally, the reduced two-qubit depth and support for fidelity-preserving chain contraction ensure that the circuits remain executable on noisy or early fault-tolerant hardware. Thus, the physical basis of QMSE greatly facilitates a good balance between interpretability, expressivity, and noise robustness.

\section{Conclusion}
\label{sec:conclusion}
In summary, we have developed QMSE as a highly effective data-encoding strategy for representing molecular structures in QML classification and regression tasks. Feature encoding techniques, such as fingerprint encoding,  suffer from poor generalisability arising from the mapping of compressed data as rotational amplitudes. In contrast, QMSE provides a straightforward means of segregating datapoints via the construction of decision boundaries, which maximises state separation in terms of the corresponding chemical moieties. 

We propose several future avenues for expanding the use of QMSE in practical quantum computing applications. The first theme focuses on broadening QMSE to encode other data structures beyond organic molecules in drug discovery, such as periodic unit cells of crystalline materials. Graph embeddings of crystal structures have demonstrated considerable success in classical machine learning workflows for accelerating materials discovery \cite{Cheng2022, An2024, Luo2024}. Hence, QMSE could enable the prediction of crystalline material properties via QML. Due to the innate ability of QMSE to load classical data linearly in the form of SMILES strings as data-encoding quantum circuits, variations of QMSE can be conceptualised that optimise the loading of other types of string information, such as encoding text as embedded tokens in quantum natural language processing \cite{Varmantchaonala2024}. Furthermore, the synthesis of QMSE circuits from the linear composition of one- and two-qubit quantum gates can be exploited in generative artificial intelligence (genAI) frameworks \cite{Lin2025},  to produce quantum circuits that can be mapped procedurally back into molecular structures.

We also seek to optimise the various QML algorithms that are compatible with QMSE. Expanding on this work for variational QML models, such as VQC and VQR, we aim to improve on ansätze by considering the evaluation of the Shapley values of their parameters, thus enhancing interpretability and synergistic effects with QMSE circuits \cite{heese2025explaining}. To combat the traditional problems associated with variational quantum algorithms, such as vanishing gradients and an abundance of local minima with poor solutions, non-variational quantum algorithms, such as quantum kernels, quantum support vector machine (QSVM) and quantum graph neural network models, can be used instead \cite{Glick2023, Lorente2022, Fuster2024, Piperno2025}. We expect this approach to be especially powerful when combined with chain contraction, allowing for the efficient evaluation of fidelities from different pairs of encoded molecular wavefunctions, and providing opportunities to perform QML tasks on more complex chemical data inventories. 

{We would also like to briefly outline our future approach in addressing the exponential concentration problem of encoding increasingly complex datapoints for higher $N$. While we have shown that QMSE greatly improves meaningful correlation between data-encoding unitaries and chemical moieties relative to other standard quantum encoding schemes such as basis, amplitude, and angle encoding, we acknowledge that a more in-depth study of overlap kernel matrices between datapoint pairs needs to be carried out to specifically address the phenomenon of exponential concentration. In a similar vein to the previous point of studying other QML models, we seek to combine QMSE circuits with data reuploading techniques for higher $L_\mathbf{x}$ that further aim to curtail exponential concentration, thus potentially enhancing their generalisation capabilities significantly \cite{Grasa2025, Perez2020}.}

Furthermore, while SMILES strings can be uniquely mapped into their respective QMSE operators, the same molecule may be represented using different SMILES strings, and hence with different QMSE circuits. Thus, it is imperative during data preprocessing to order the atoms of the SMILES strings of the molecular dataset consistently, such that the optimal comparison between datapoints is maximised. {Other potential schemes to address this shortcoming could be the randomisation of choosing a SMILES string representing each molecule in the dataset to reduce selection bias, the inclusion of more datapoints of different SMILES strings mapped to the same molecule for better training alignment, or the conversion of more sophisticated more string representations such as the International Chemical Identifier (InChI) \cite{Krenn2022} into QMSE data-encoding circuits. We aim to examine this problem more systematically with curated databases in the future, such as subsets of GDB-17 that feature interesting organic scaffolds for potential drug discovery applications \cite{Ruddigkeit2012}.}

Finally, we consider the position of QMSE as an effective data-encoding method in the context of the ongoing transition from near-term to early FTQC regimes. In the early fault‐tolerant regime, error‐corrected logical qubits enable high‐fidelity preparation of molecular graph‐state encodings via block‐encoding of adjacency or Coulomb–adjacency matrices, directly mapping connectivity into entanglement patterns \cite{Hein2004}. Reduced noise and parallelisable CZ‐based graph‐state circuits allow deeper variational ansätze without barren plateaux, improving gradient magnitudes and convergence \cite{cerezo2023does, Preskill2018}. {The usage of 
partial quantum error correction (pQEC) methods to prepare high-quality, arbitrary $R_z$ rotations gates that are outside of the Clifford gate set via magic state injection also allow larger variational quantum circuits to be implemented \cite{Dangwal2025}. Another example involves the parallelisation of evaluating separate expectation values of individual datapoints \cite{Cattelan2025, Tsukayama2025}, thus further tightening integration between quantum devices and high-performing computers in addition to leveraging the latter for QEC methods.} The preparation of fault-tolerant graph states also reduces depth overhead by commutativity, enhancing the resilience to residual errors \cite{Gottesman1997, huang2024quantum}. Consequently, QML models based on explicit graph‐state encodings are expected to exhibit faster convergence and better generalisation on early FTQC hardware compared to their pre-FTQC counterparts \cite{Marshall2023highdimensional}. Overall, QMSE is expected to benefit significantly from early FTQC frameworks, regardless of combination with variational or non-variational QML models.

\section*{Data and code availability}
We provide a publicly accessible GitHub repository (\href{https://github.com/stfc/quantum-molecular-encodings}{github.com/stfc/quantum-molecular-encodings}) hosting the routines for mapping SMILES strings to hybrid Coulomb–adjacency matrices, as well as generation of their corresponding data-encoding quantum circuits. The repository also contains examples formatted as Jupyter notebooks. We also provide the 105-molecule dataset from the 95\textsuperscript{th} CRC Handbook of Chemistry and Physics \cite{haynes2014crc} with canonical SMILES and normalised bond-order matrices. The implementation of the molecular structure encoding layer introduced in this work will be made available in the main Qiskit Machine Learning library \cite{QiskitML2025} from version 0.9.
The code and data produced by this work are distributed under the (CC BY) license without any warranty.

\section*{Author contributions}
We provide the author contributions following the CRediT (Contributor Roles Taxonomy) scheme.
CB and DJW conceptualised the molecular encoding scheme. CB and EA contributed equally to developing the QML methodology, software, formal analysis, investigation and writing of the original draft of the manuscript. EA produced the figures and integrated the QMSE framework with Qiskit Machine Learning. DM and CB compiled the datasets and performed the data curation. IT oversaw project administration and supervision. SM was responsible for funding acquisition and project administration. All the authors have reviewed and provided feedback on the final manuscript.

\begin{acknowledgments}
    EA and CB are grateful to M. Emre Sahin for helping to draft the code that inputs quantum circuits as feature maps in Qiskit Machine Learning. We acknowledge helpful conversations with Jason Crain and help from Edward O. Pyzer-Knapp and Benjamin C. B. Symons during the initial set-up of the project.
    This work was supported by the Hartree National Centre for Digital Innovation, a UK Government-funded collaboration between STFC and IBM.
    IBM, the IBM logo, and \href{https://ibm.com}{www.ibm.com} are trademarks of International Business Machines Corp., registered in many jurisdictions worldwide. Other product and service names might be trademarks of IBM or other companies. The current list of IBM trademarks is available at \href{https://www.ibm.com/legal/copytrade}{www.ibm.com/legal/copytrade}.

    
\end{acknowledgments}

\bibliography{main}
\clearpage

\onecolumngrid
\appendix


\section{Simulating quantum fidelities of fatty acids}
\label{app:fatty-acids}
\vspace{-0.3cm}
\begin{figure*}[ht!]
    \centering
    \includegraphics[width=0.9\textwidth]{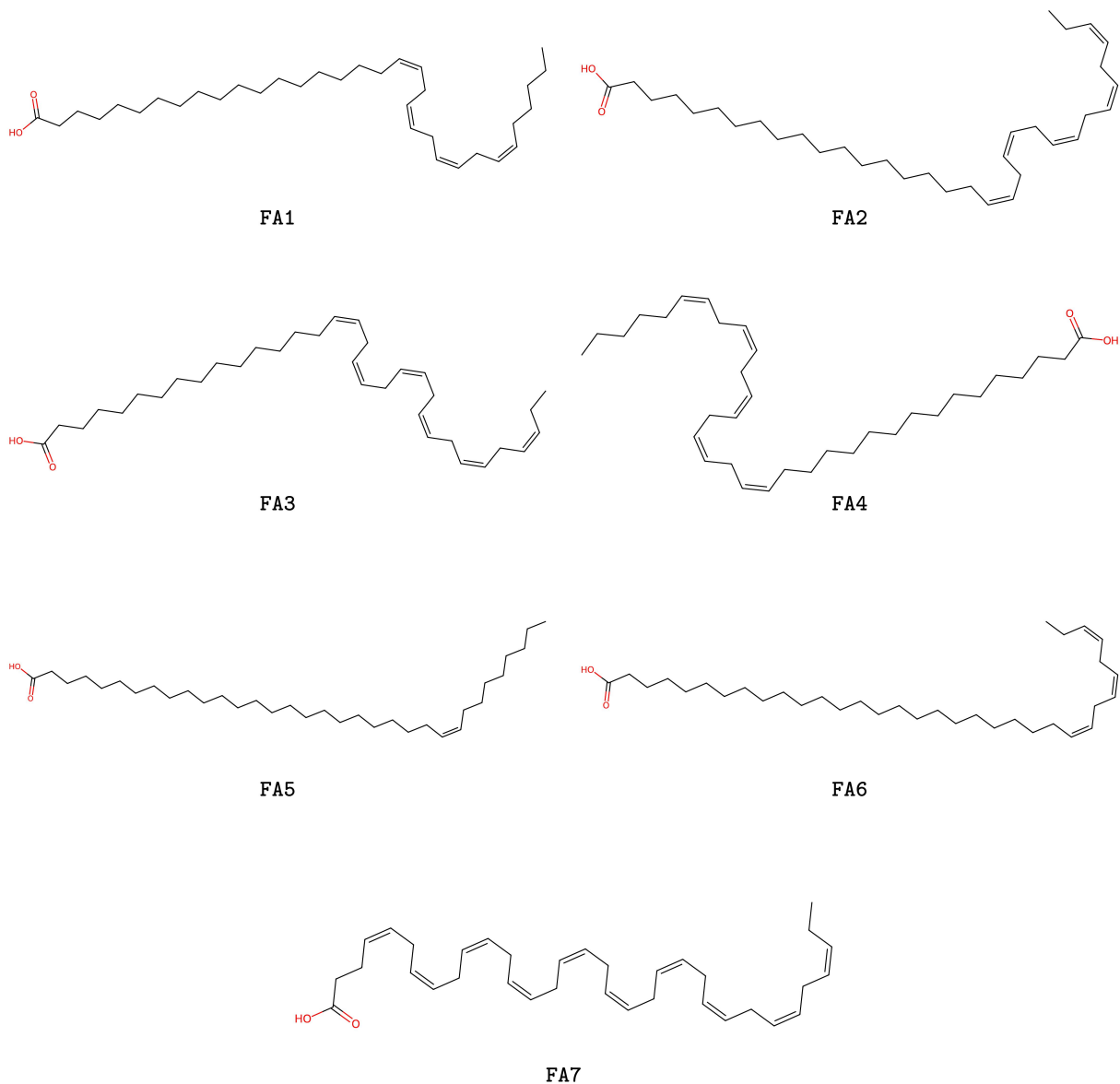}
    \caption{Skeletal structures of the fatty acid series \textbf{FA1 - FA7} as outlined in Table~\ref{tab:fattyacids}.}
    \label{fig:overlap:FA-main}
\end{figure*}

\begin{table}[ht]
\small
\centering
\caption{Summary of the identities and SMILES representation orderings of the fatty acid series $\textbf{FA1$-$FA7}$.} 
\label{tab:fattyacids}
\begin{tabularx}{\textwidth}{p{0.25in}p{0.95in}p{2.2in}p{3.4in}}
\toprule
\textbf{ID}\rule{0pt}{2.5ex}  & \textbf{PubChem CID} & \textbf{IUPAC Name}                                               & \textbf{SMILES Representation}                                                                                                                                                                                                          \\ \addlinespace[0.5ex]
\midrule
\addlinespace[0.5ex]
\textbf{FA1} & 52921804             & 19\textit{Z},22\textit{Z},25\textit{Z},28\textit{Z}-tetratriacontatetraenoic acid                     & OC(=O)CCCCCCCCCCCCCC
CCC/C=C\textbackslash{}C/C=C\textbackslash{}C/C=C\textbackslash{}C/C=C\textbackslash{}CCCCC \\
\textbf{FA2} & 14753668 & 19\textit{Z},22\textit{Z},25\textit{Z},28\textit{Z},31\textit{Z}-tetratriacontapentaenoic acid                 & OC(=O)CCCCCCCCCCCCCC
CCC/C=C\textbackslash{}C/C=C\textbackslash{}C/C=C\textbackslash{}C/C=C\textbackslash{}C/C=C\textbackslash{}CC                                                                                           \\ 
\textbf{FA3} & 52921817             & 16\textit{Z},19\textit{Z},22\textit{Z},25\textit{Z},28\textit{Z},31\textit{Z}-tetratriacontahexaenoic acid              & OC(=O)CCCCCCCCCCCCCC
/C=C\textbackslash{}C/C=C\textbackslash{}C/C=C\textbackslash{}C/C=C\textbackslash{}C/C=C\textbackslash{}C/C=C\textbackslash{}CC                                                                         \\ 
\textbf{FA4} & 52921824             & 16\textit{Z},19\textit{Z},22\textit{Z},25\textit{Z},28\textit{Z}-tetratriacontapentaenoic acid                 & OC(=O)CCCCCCCCCCCCCC
/C=C\textbackslash{}C/C=C\textbackslash{}C/C=C\textbackslash{}C/C=C\textbackslash{}C/C=C\textbackslash{}CCCCC                                                                                           \\ 
\textbf{FA5} & 92033288             & 25\textit{Z}-tetratriacontenoic acid                                       & OC(=O)CCCCCCCCCCCCCC
CCCCCCCCC/C=C\textbackslash{}CCCCCCCC                                                                                                                                                                   \\ 
\textbf{FA6} & 171118569            & 25\textit{Z},28\textit{Z},31\textit{Z}-tetratriacontatrienoic acid                           & OC(=O)CCCCCCCCCCCCCC
CCCCCCCCC/C=C\textbackslash{}C/C=C\textbackslash{}C/C=C\textbackslash{}CC                                                                                                                               \\ 
\textbf{FA7} & 171117702            & 4\textit{Z},7\textit{Z},10\textit{Z},13\textit{Z},16\textit{Z},19\textit{Z},22\textit{Z},25\textit{Z},
28\textit{Z},31\textit{Z}-tetratriacontadecenoic acid & OC(=O)CC/C=C\textbackslash{}C/C=C\textbackslash{}C/C=C\textbackslash{}C/C=C\textbackslash{}C
/C=C\textbackslash{}C/C=C\textbackslash{}C/C=C\textbackslash{}C/C=C\textbackslash{}C/C=C\textbackslash{}C/C=C\textbackslash{}CC \\ \bottomrule
\end{tabularx}
\end{table}

The procedure for simulating the quantum fidelities of the unsaturated fatty acid series \textbf{FA1 $-$ FA7} is outlined in this section. The molecules and  structures are described in Table~\ref{tab:fattyacids} and Fig.~\ref{fig:overlap:FA-main}, respectively. The SMILES string representations of the fatty acid series were first canonicalised via RDKit and subsequently reordered with the carboxylic acid moieties left-aligned, so as to ensure maximum structural overlap between the molecules. Using the chain contraction procedure, the maximum overlap between pairs of SMILES strings was omitted and the quantum fidelity circuit was constructed for the molecule pair via QMSE with $R_y$ and $R_{xx}$ as the rotational and entangling gates, respectively, and $L_{\mathbf{x}}=1$. The number of qubits required to construct the circuit was determined by the length of the longer reduced SMILES string. As the fatty acids display geometric isomerism from the C$=$C double bonds in the \textit{Z} configuration, the optional argument $\epsilon_T$ was imposed on the required $R_y$ one-qubit rotations. The resulting fidelity values are shown in Fig.~\ref{fig:overlap:fattyacids}.
\section{Loss curves of alkane subdataset VQC model}
\label{app:loss-functions}
\vspace{-0.5cm}
\begin{figure}[H]
    \centering
    \includegraphics[width=\textwidth]{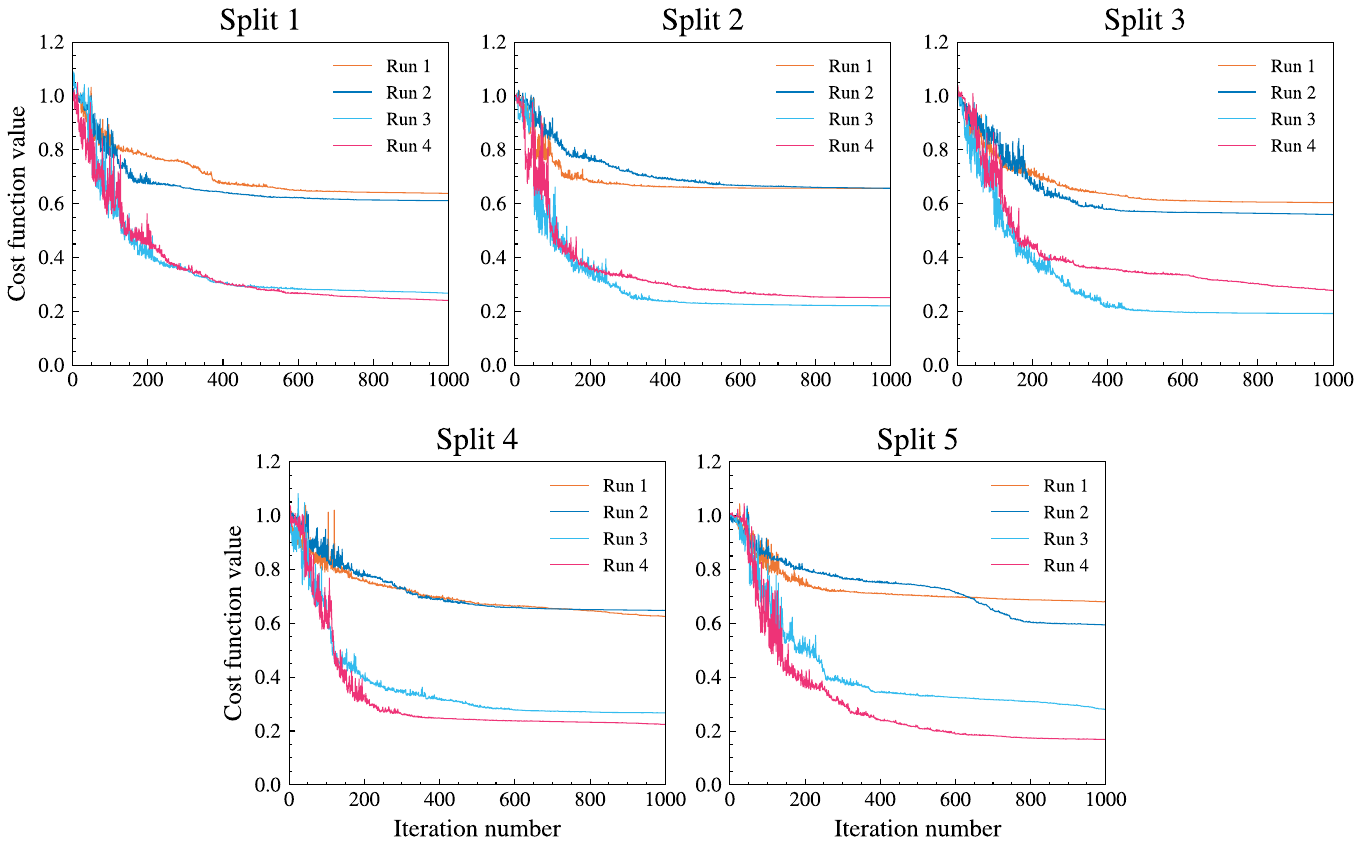}
    \caption{Training $L_2$ loss curves on the VQC model for Runs 1-4 on the alkane subdataset and their respective $k$-fold splits, with random ansatz parameter initialisation and $L_{\theta}=5$.}
    \label{fig:loss_curves}
\end{figure}
Fig.~\ref{fig:loss_curves} shows the training $L_2$ loss curves for the VQC model of the alkane subdataset tasks for Runs 1-4 and different $k$-fold splits. In terms of the ansatz, random parameter initialisation with similar losses were selected in this illustration for all runs with $L_\theta=5$. We found that all five $k$-fold splits behave similarly, and QMSE (Runs 3-4) exhibits superior convergence in model performance compared to fingerprint encoding (Runs 1-2), consistent with the improvement in training accuracies as shown in Fig.~\ref{fig:classification-scores}.

\section{Forecasts for near-term devices}

\begin{figure*}[h]
    \centering
    \includegraphics[width=\linewidth]{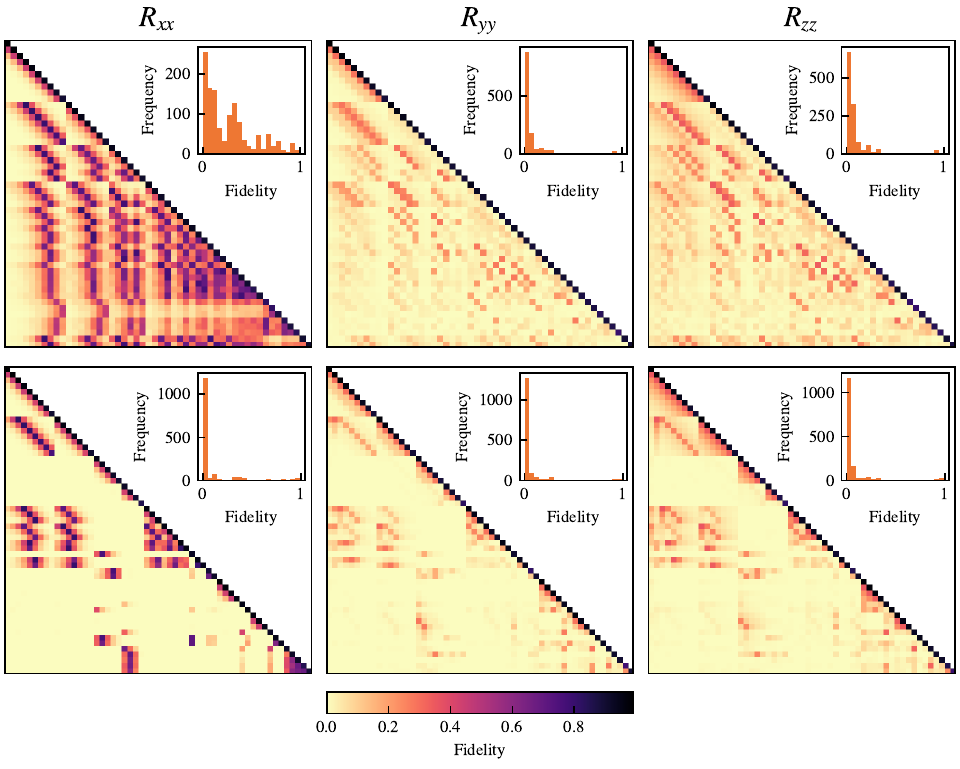}
    \caption{Heatmaps of fidelity matrices for chain-contracted molecules encoded with QMSE within the alkane- (top row) and oxygen- (bottom row) subdatasets, calculated in the presence of device noise. The noise profile corresponds to data for the \texttt{ibm\_pittsburgh} device.}
    \label{fig:overlap-noisy}
\end{figure*}

{
As shown in section~\ref{sec:datasets}, the strength of QMSE as an encoding scheme for QML stems from its ability to discriminate molecules of similar type. To evaluate the utility of executing QMSE on currently available quantum hardware, we performed fidelity measures for the alkane- and oxygen-subdatasets using noise profiles designed to mimic the errors present in quantum processors. The quantum circuits constructed via Eq. \ref{eq:unitary} were transpiled to \texttt{ibm\_pittsburgh} device, an IBM quantum chip built using superconducting transmon qubits arranged in a heavy-hexagonal topology. A representative noise model for this device was selected based on calibration data recorded on the 10th of August 2025 at 18:59:40 UTC. This model was incorporated to sampling of the fidelity measurements via IBM's Qiskit Aer simulator. Each fidelity measure was sampled over 10,000 shots. Heatmaps of quantum fidelity matrices for these noisy simulations are shown in Fig.~\ref{fig:overlap-noisy}.
}

{
The presence of noise is expected to lead to a degradation of the main diagonal elements in the heatmap matrix compared to the noiseless heatmaps of Fig.~\ref{fig:overlap}. However, we note that the difference in the fidelities between the noiseless and the noisy cases is minute, which can be mainly attributed to the shallow entangling gate depths of the circuits produced by the QMSE scheme enhanced with chain contraction.  
}

\end{document}